# The Microwave Hall Effect Measured Using a Waveguide Tee

J. E. COPPOCK[1], J. R. ANDERSON[1], and W. B. JOHNSON[2]

[1]Dept. of Physics, University of Maryland, College Park, MD, 20741
[2]Laboratory for Physical Sciences, College Park, MD, 20740

This paper describes a simple microwave apparatus to measure the Hall effect in semiconductor wafers. The advantage of this technique is that it does not require contacts on the sample or the use of a resonant cavity. Our method consists of placing the semiconductor wafer into a slot cut in an X-band (8 - 12 GHz) waveguide series tee, injecting microwave power into the two opposite arms of the tee, and measuring the microwave output at the third arm.  A magnetic field applied perpendicular to the wafer gives a microwave Hall signal that is linear in the magnetic field and which reverses phase when the magnetic field is reversed. The microwave Hall signal is proportional to the semiconductor mobility, which we compare for calibration purposes with d. c. mobility measurements obtained using the van der Pauw method.  We obtain the resistivity by measuring the microwave reflection coefficient of the sample.  This paper presents data for silicon and germanium samples doped with boron or phosphorus. The measured mobilities ranged from 270 to 3000 cm$^2$ /(V-sec).

1. Introduction

A. Motivation

Direct current electrical techniques for measuring the mobility and carrier concentration in semiconductors require making contacts that are mechanically reliable and electrically non-rectifying.  The van der Pauw method is a standard d. c. technique for measuring the mobility and carrier concentration in a semiconductor wafer as long as the size of each contact is small [1].  Alternative techniques that do not require contacts include the microwave Hall effect: viz., if a linearly polarized microwave is incident on a sample with an external magnetic field applied perpendicular to the incident polarization, then the Hall effect produces an electric field that is orthogonal to the incident polarization.  Microwave techniques allow the measurement of the orthogonal electric field. The orthogonal electric field is proportional to the mobility of the sample.

Typical methods for measuring the microwave Hall effect either use a resonant microwave cavity with two degenerate orthogonal cavity modes [2,3] or rely on the Faraday rotation of a linearly polarized  microwave that passes through the material [4].  Both methods rely on the Hall effect to generate an electric field that is orthogonal to the applied d. c. magnetic field and the incident microwave electric field.  There are commercial instruments that use these microwave techniques for mobility measurements [5].   In this paper, we study the use of a slotted microwave series tee and a microwave bridge for making the mobility measurement.   A similar technique to ours uses two crossed waveguides to separate incident microwaves from outgoing microwaves produced by the Hall effect [6].

There are several advantages to using a slotted microwave tee compared to using a cavity or Faraday rotation. First, when a microwave cavity is used, it needs to be custom designed and fabricated for specific resonant frequencies. Microwave tees, however, are commercially available and are broadband. Second, when Faraday rotation is used, a large applied microwave electric field must be canceled or a**bs**orbed to allow measurement of a small orthogonal field. In a microwave tee, the output from one arm of the tee can be arranged to be zero when there is no applied magnetic field by applying microwaves of equal magnitude and phase to the other two arms, as we explain in section 2B.

One disadvantage of microwave measurements compared to the d. c. van der Pauw method is that microwave signal generators and network analyzers are more expensive than the electrometers that are used in d. c. van der Pauw measurements. Another disadvantage is that the microwave series tee method has to be calibrated on samples with known mobilities since the microwave Hall signal depends on the geometry, i.e., on the shape of the sample and its position in the microwave tee. We used the d. c. van der Pauw method to calibrate the microwave Hall effect and this paper reports the results of the calibration.

B. History

Measurements of the Faraday rotation and microwave Hall effect started in the 1940's and expanded in the 1950's. There was a large amount of work during WWII on microwave radar and this led to the extensive development of microwave equipment and related techniques [7]. After WWII, these techniques were applied to materials characterization in a surge of activity that has continued to the present day. Early sources of microwaves were klystrons and magnetrons. The main transmission method for microwaves used metal waveguides rather than coaxial wires because the attenuation was smaller and the power handling capacity was larger in waveguides. Current technology provides tunable solid state sources that produce single frequency coherent microwaves at high powers. Coaxial cables are now much more widely used than metal waveguides, but in this experiment we use X band waveguides because we are able to cut a slot to insert a sample into a waveguide series tee.

There are two main methods for measuring the microwave Hall effect. In Faraday rotation, the applied magnetic field is parallel to the direction of propagation of the microwaves and one measures the polarization rotation of a linearly polarized microwave incident on a sample [4]. Faraday rotation is large in the ferrite materials that were developed for use in isolators and circulators. In semiconductors, however, the Faraday rotation is usually much smaller than in ferrites. This means that in a Faraday rotation measurement, the polarization of the incident microwave is only slightly rotated by an applied d. c. magnetic field. Measuring the rotation then requires canceling a large incident field to observe the small orthogonal electric field. One way to do this is to use crossed waveguides [8,9] where one waveguide supports only one polarization and the other supports only the orthogonal polarization. Moreover, Faraday rotation requires the magnetic field to run parallel to the propagation direction in the waveguide and this usually means the waveguide is in a solenoid to obtain the correct geometry.

The other method for measuring the microwave Hall effect is to put the sample in a custom microwave cavity that supports two orthogonal modes that are degenerate, i.e., both have the same resonance frequency. The sample is placed in the cavity at a position where the electric field is a maximum (antinode) of one mode. The microwave field in the cavity for each mode is a standing wave that is the superposition of two waves with opposite k-vectors. The d. c. magnetic field is applied parallel to those k-vectors as in Faraday rotation. The mobility is proportional to the signal induced in the orthogonal mode [10,11]. A useful reference for the various microwave techniques used to characterize materials is [12].

This paper discusses a third method, where the Hall signal is induced in a sample wafer in one waveguide and then couples to a second waveguide that has a polarization orthogonal to the first waveguide [6].

In this paper we also describe a microwave technique to measure the resistivity. Microwave resistivity techniques in other studies include measuring the attenuation and reflection of material placed in waveguides and transmission lines and the change in the quality factor in resonant cavities [12]. We discuss a method for measuring resistivity that involves measuring the amount of reflected power from a sample in a waveguide with a variable shorting stub [13,14,15,16]. These studies also fit the variation of the resistivity with a magnetic field to a quadratic Drude model and can derive the mobility if there is only one carrier [13]. Since both the mobility and the resistivity can be obtained from microwave measurements, the carrier concentration can be obtained as well.

The remainder of the paper is divided into part 2, which discusses the microwave Hall effect measurements, and part 3, which discusses the microwave resistivity measurements.

## 2. Microwave Hall Effect

### A. Theory

In this section, we derive the basic relation between the incident microwave field $E_{inc}$ and the orthogonal outgoing microwave field $E_{out}$. The result is the same relation as for the d. c. Hall effect, viz.,

$$\frac{E_{out}}{E_{inc}} \propto \mu B \quad (2.1)$$

where μ is the mobility in the sample and $B$ is the applied d.c. magnetic field.

In the analysis, we start from the simple Drude model for an electron that undergoes scattering in the presence of both an electric and magnetic field [10]:

$$m\frac{d\boldsymbol{v}}{dt} + m\frac{\boldsymbol{v}}{\tau} = q\boldsymbol{E} + q\,\boldsymbol{v}\times\boldsymbol{B} \quad (2.2)$$

where $m$ is the mass of the carrier and $q$ is its charge, and $\mathbf{v}$ is the velocity of the carrier.

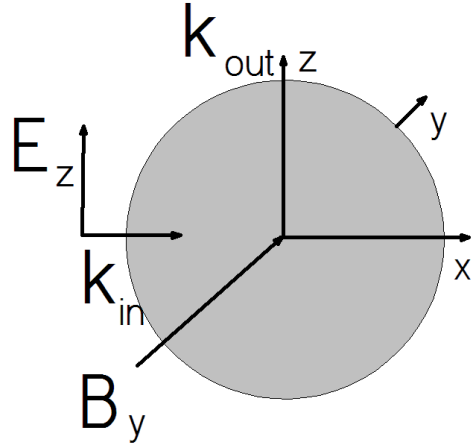

Figure 1. Geometry of the microwave Hall effect. An incident microwave field of amplitude $E_z$ and wave vector $k_{in}$ propagates through a semiconductor wafer in the x-z plane. This microwave field induces currents in the wafer in the z direction. In the presence of a static magnetic field $B_y$, a Lorentz force in the x direction produces an outgoing microwave Hall field with wave vector $k_{out}$.

Assume that the incident microwave is traveling along the x-axis with amplitude $E_{0z}$ that is polarized along the z-axis. Also assume that the applied d. c. magnetic field B is along the y-axis (fig. 1). The microwave electric field $E_z$ induces a sinusoidal velocity $v_z$ in the carriers that is along the z-axis and has amplitude $v_{0z}$. The Lorentz force will act along the x-axis and cause the carriers to move along the x-axis with velocity $v_x$ and amplitude $v_{0x}$, and this will produce an outgoing microwave electric field $E_x$ traveling along the z-axis with amplitude $E_{0x}$. This is the microwave Hall effect. We also assume that the time variation is harmonic with angular frequency ω:

$$E_x = E_{0x}e^{-i\omega t}\hat{x} \text{ and } E_z = E_{0z}e^{-i\omega t}\hat{z} \quad (2.3)$$

$$v_x = v_{0x}e^{-i\omega t}\hat{x} \text{ and } v_z = v_{0z}e^{-i\omega t}\hat{z} \quad (2.4)$$

Writing out the components of eq. 2.2 gives:

$$-im\omega v_{0x} + m\frac{v_{0x}}{\tau} = qE_{0x} - qv_{0z}B_y \quad (2.5)$$

$$-im\omega v_{0z} + m\frac{v_{0z}}{\tau} = qE_{0z} + qv_{0x}B_y \quad (2.6)$$

Solving for the velocity component along the x-axis gives:

$$v_{0x} + \frac{q^2 B_y^2 v_{0x}}{(-im\omega + \frac{m}{\tau})^2} = \frac{qE_{0x}}{(-im\omega + \frac{m}{\tau})} - \frac{q^2 B_y E_{0z}}{\left(-im\omega + \frac{m}{\tau}\right)^2} \quad (2.7)$$

In the Drude model, the mobility $\mu = \frac{q\tau}{m}$, and eq. 2.7 can be written as:

$$v_{0x} + \frac{(\mu B)^2 v_{0x}}{(-i\omega\tau + 1)^2} = \frac{\mu E_{0x}}{(-i\tau\omega + 1)} - \frac{\mu B_y \mu E_{0z}}{(-i\omega\tau + 1)^2} \quad (2.8)$$

We are trying to calculate the first order effect of the incident wave on the velocity of a carrier. For the incident wave, $E_{0x} = 0$. For the materials that we measure, $\mu B < 1$ and $\omega\tau \ll 1$, so we can throw away terms to reduce eq. 2.8 to:

$$v_{0x} = -\mu B_y \mu E_{0z} \quad (2.9)$$

Now we need to use Maxwell's equations to find the outgoing microwave field:

$$\text{Faraday's law: } \nabla \times \boldsymbol{E} = i\omega \boldsymbol{B} \quad (2.10)$$

$$\text{Ampere's law: } \nabla \times \boldsymbol{B} = \mu_0(nq\boldsymbol{v}) - \mu_0 i\omega\varepsilon\boldsymbol{E} \quad (2.11)$$

Combining these equations gives:

$$(\nabla^2 + \omega^2 \epsilon \mu_0)\mathbf{E} = -i\omega\mu_0 nq\mathbf{v} + \nabla(\nabla \cdot \mathbf{E}) \quad (2.12)$$

where $\mu_0$ is the permeability of the vacuum and $\varepsilon$ is the dielectric constant in the material, and the carrier concentration is $n$. The material is assumed to be non-magnetic.

We have assumed that there are two polarized waves in the sample, an incident microwave traveling along the x-axis and polarized along the z-axis with wave number k:

$$E_{inc} = E_{0z} e^{i(kx-\omega t)} \hat{z} \quad (2.13)$$

and an outgoing microwave traveling along the z-axis and polarized along the x-axis with wave number k:

$$E_{out} = E_{0x} e^{i(kz-\omega t)} \hat{x} \quad (2.14)$$

Note that we suppressed the k dependence of the amplitudes $E_{0x}$ and $E_{0z}$ in eq. 2.3 and now make it explicit in eq. 2.13 and eq. 2.14. Since the waves are transverse, $\nabla \cdot E = k \cdot E = 0$. From Gauss's law, this just means there is charge neutrality in the material. Substituting $E = E_{inc} + E_{out}$ into the partial differential eq. 2.12 and using eq. 2.9 gives two equations, one for each axis:

$$(-k^2 + \omega^2 \varepsilon \mu_0) E_{0z} = -i\omega\mu_0 nq v_z = -i\omega\mu_0 nq\mu E_{0z} \quad (2.15)$$
$$(-k^2 + \omega^2 \varepsilon \mu_0) E_{0x} = -i\omega\mu_0 nq v_x = i\omega\mu_0 nq(\mu B_y)(\mu E_{0z}) \quad (2.16)$$

The first equation gives the value for k (where the conductivity $\sigma = nq\mu$):

$$(-k^2 + \omega^2 \varepsilon \mu_0) + i\omega\mu_0 \sigma = 0, \quad (2.17)$$

and the second gives the value for the outgoing wave $E_{ox}$:

$$E_{0x} = \frac{i\omega\mu_0 nq(\mu B_y)(\mu E_{0z})}{(-k^2 + \omega^2 \varepsilon \mu)} = \frac{i\omega\mu_0 nq(\mu B_y)(\mu E_{0z})}{-i\omega\mu_0 nq\mu}$$
$$= -(\mu B_y)E_{0z} \quad (2.18)$$

The output microwave electric field $E_{0x}$ is thus proportional to the mobility and the d. c. magnetic field $B_y$. Eq. 2.18 allows the measurement of the mobility in terms of the easily measurable input and output microwave power and the applied d. c. magnetic field.

Inevitable microwave power losses while coupling between the wafer and the waveguide, as well as internal losses within the wafer, mean that the ratio $E_{0x}/B_y E_{0z}$ will be less than the true mobility. To determine this loss, we define $\mu_{MW} = \frac{E_{0x}}{BE_{0z}}$ as the "microwave mobility" and experimentally measure a calibration ratio $\kappa$ :

$$\kappa = \frac{\mu_{DC}}{\mu_{MW}} \quad (2.19)$$

where the mobility $\mu_{DC}$ is determined by a standard 4-point d.c. van der Pauw measurement [1].

B. Experimental Setup and Measurement Procedure

The relationship $E_{0x} = \mu B E_{0z}$ implies that we can determine the mobility by measuring the output microwave Hall field $E_{0x}$ induced by an input microwave field $E_{0z}$ in the presence of an applied d. c. magnetic field $B$. The microwave Hall apparatus (shown schematically in fig. 2) consists of an X-band waveguide series tee with a slot cut over the junction. The wafer to be measured is inserted in the slot, centered and held in place by nonconductive Delrin blocks on either side.

Microwaves of a single frequency (approximately 10 GHz) are generated by a network analyzer and injected into the two opposite input arms of the series tee.  At 10 GHz, the single mode in the waveguide is $TE_{10}$, where the input electric field $E_{0z}$ is confined parallel to the shorter dimension of the waveguide and to the plane of the wafer.  A d. c. magnetic field is applied normal to the plane of the wafer and produces a microwave Hall field that lies in the plane of the wafer and is perpendicular to the input microwave field.  The microwave Hall field couples to the third arm (the output arm) and is measured at the second port of the network analyzer.

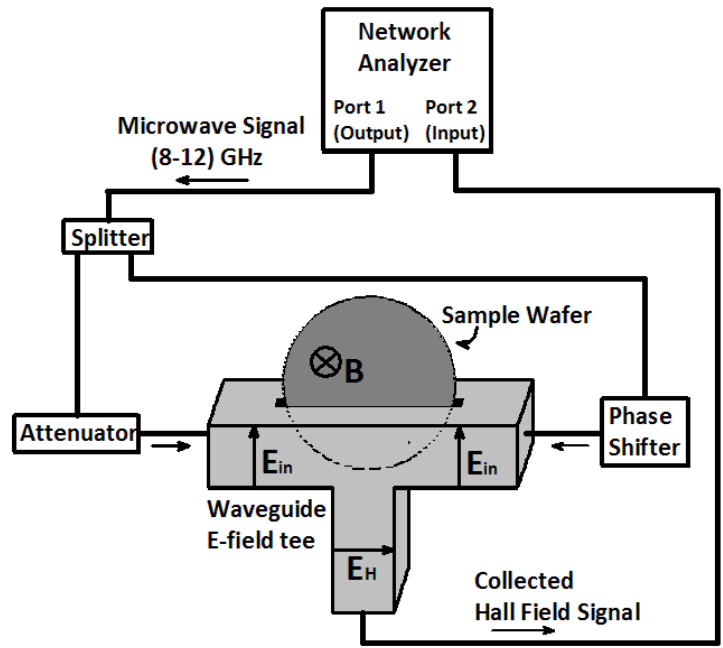

Figure 2. Schematic diagram of the microwave Hall apparatus.

The following procedure is used to minimize the amount of input power that couples to the third arm.  This procedure removes any signal independent of the magnetic field and enables us to measure Hall signals in the third arm that are 60 dB smaller than the input power level. It is a property of waveguide series tees that if two signals of equal amplitude and matched phase are injected into the opposite arms of the tee, they cancel in the third arm and thus no output is observed from the third arm. This property depends only on the bilateral symmetry of the tee and is also why the series tee can more typically be used as a power divider which splits power going into the third arm into microwaves of equal amplitude and opposite phase in the two opposite arms of the tee. In our apparatus, we use an attenuator at one input arm and a phase shifter at the other to tune the inputs until a minimum in output power is observed. The nulling is done when the d. c. magnetic field is zero and the sample has been placed in the slot cut into the tee.

Microwave Hall data is obtained from the $S_{21}$ scattering parameter of a network analyzer, which in fig. 2 is the ratio $E_H/E_{in}$ of the output microwave field $E_H$ to the input microwave field $E_{in}$. The amplitude of the Hall field is given by:

$$E_H = (magnitude\ of\ S_{21}) \times \cos(phase\ of\ S_{21}) \quad (2.20)$$

The d.c. magnetic field is swept through both positive and negative magnetic field directions. The phase in eq. 2.20 is arbitrarily defined as zero for the positive magnetic field direction and is typically 180° for the negative d.c. magnetic field direction. The 180° phase shift is the result of the Hall voltage changing sign when the magnetic field direction changes sign.

After Hall data is taken, the amplitudes of the input signals to each waveguide arm are measured. $E_H$ is the average of the two input amplitudes. Typical microwave Hall data is shown in fig. 3. Data points near B=0 are not shown because as the signal goes to zero, it becomes smaller than the noise.

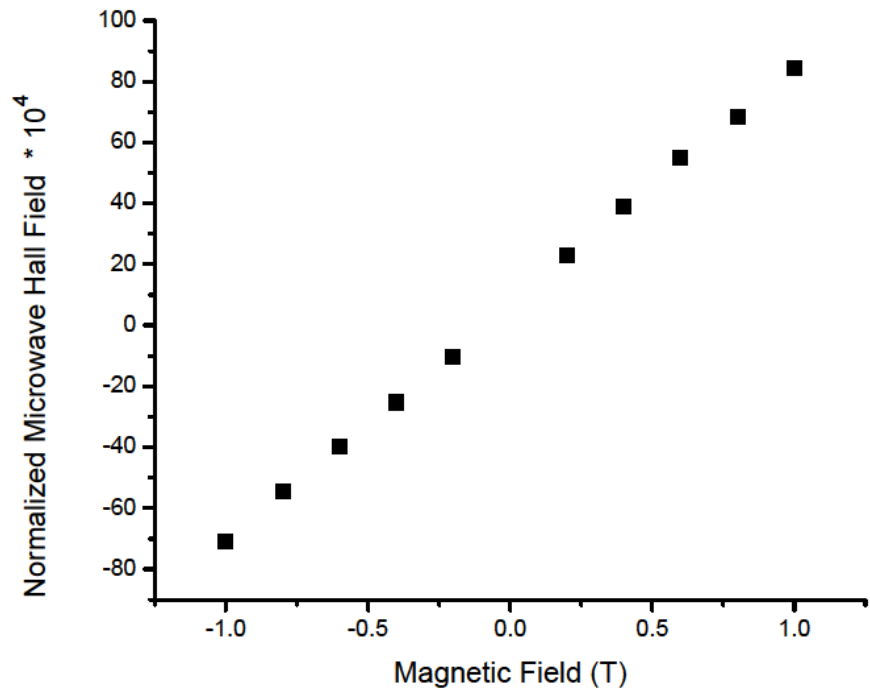

Figure 3. The magnitude of the Hall field $E_H$ normalized by dividing by the input microwave field $E_{in}$. The Hall field increases linearly with d. c. magnetic field B. The slope of the line is defined to be $\mu_{MW} B$, where $\mu_{MW}$ is the microwave mobility. Measuring the slope of the line is how we determine the microwave mobility.

We used a WR90 waveguide [17] that has single mode operation from 8.20-12.40 GHz. The arms of the waveguide tee extend outside the d. c. magnetic field to minimize the effect of magnetic components in cables and connectors moving when the magnetic field changes and thereby contributing spurious signals to the output of the waveguide tee.

The applied d. c. magnetic field is provided by an electromagnet with a maximum field of approximately 1 Tesla. The magnetic field strength is monitored with a d. c. Hall probe.

### C. Measurements

By measuring both the microwave mobility and the d. c. mobility of a sample, we determine the calibration factor κ, defined in eq. 2.19. We measured the microwave mobility for sets of two and three inch diameter Si and Ge wafers purchased from University Wafers [18]. The Si wafers come in a pack of up to a dozen wafers. The wafers in a set are probably all from the same boule and therefore have the same nominal carrier concentration and mobility. We verified that this was the case by measuring the d. c. mobility and resistivity using the van der Pauw method. We combined the wafers from a given set to double or triple the thickness of the wafers. We also varied the operating frequency of the microwaves. We find empirically that κ depends on both the thickness $T$ and skin depth of the material measured. The skin depth is given by [19]:

$$\delta = 1 \Big/ \sqrt{\pi \sigma f \mu_0} \qquad (2.21)$$

where $f$ is the frequency of the microwaves, σ is the conductivity and $\mu_0$ is the permeability of the vacuum. The material is assumed to be non-magnetic. For the circular wafers, the relationship between κ and $T/\delta$ is approximately linear (see fig. 4).

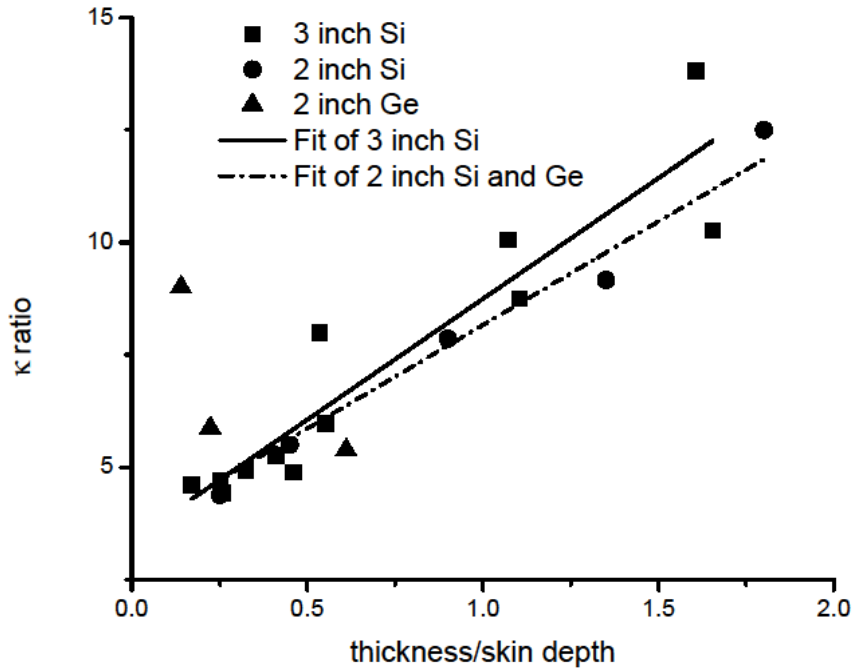

Figure 4. Variation of $\kappa = \frac{\mu_{DC}}{\mu_{MW}}$ vs. thickness/skin depth for two and three inch wafers. $\mu_{DC}$ is the mobility determined by the d. c. van der Pauw method and $\mu_{MW}$ is the mobility determined by the microwave method discussed in this paper.

    An outlier in the data in fig. 4 is a germanium wafer with a resistivity of 34.5 ohm-cm, and κ of 9 and T/δ of .2 which is considerably higher than the resistivity of the other Si and Ge wafers, which are all below 14 ohm-cm. The least squares fit for the three inch silicon wafers gives $\kappa = 5.4(T/\delta) + 3.4$. Excluding the high resistivity Ge data point, fitting the two inch wafers gives $\kappa = 4.6(T/\delta) + 3.6$. These equations can be used to determine the mobility of a circular wafer whose resistivity is in the range 1.5-13.9 ohm-cm. If the uncertainty in the microwave Hall data $E_H/E_{in}B$ is about 2% and the uncertainty in $\kappa$ is about 10%, the mobility of a circular wafer whose

resistivity is in the range 1.5-13.9 ohm-cm can be determined to about 10%.

For rectangular samples, the behavior of κ is different.  The largest rectangle that we could make from a 3-inch wafer has an area of about 45 cm² and the smallest that we could cleave and handle is about 0.2 cm². Fig. 5 shows a plot of the microwave Hall mobility $\mu_{MW} = \frac{E_H}{E_{in}B}$ against the area of the wafer that is contained within the waveguide. Removing part of the wafer outside the waveguide tee does not change the microwave Hall signal, since Hall fields are not induced in that part of the sample.

When the part of the sample that sits inside the waveguide is a semicircle, the result is significantly different from when the part of the sample inside the waveguide is approximately rectangular.  This is due to the distribution of material inside the waveguide.  For a semicircle, the absorbing edge is along the circumference of the wafer and part lies directly over the output arm of the waveguide, whereas for a rectangle of the same area, the absorbing edges are on the sides of the wafer and lie farther from the output arm.  More of the Hall fields will couple to the output arm for the semicircle than the rectangle.  For example, the measured microwave mobility for a three inch semicircular Si sample with d. c. mobility of 1400 was 269 $\frac{cm^2}{V-s}$. The area inside the waveguide was 3.51 cm². When the wafer was cleaved to approximately a rectangle of 3.47 cm², the measured mobility decreased to 178 $\frac{cm^2}{V-s}$.

Fig. 5 shows that as a rectangular sample decreases in area, its Hall signal increases until a turning point around 0.5 cm², after which it falls rapidly.  Again, this may be explained by the fact that smaller rectangles have their absorbing edges positioned closer to the opening of the output arm, enabling more Hall fields to couple to the output arm. As the sample becomes smaller than 0.5 cm², decreasing its area only decreases the area in which Hall fields can be generated and the Hall signal decreases.

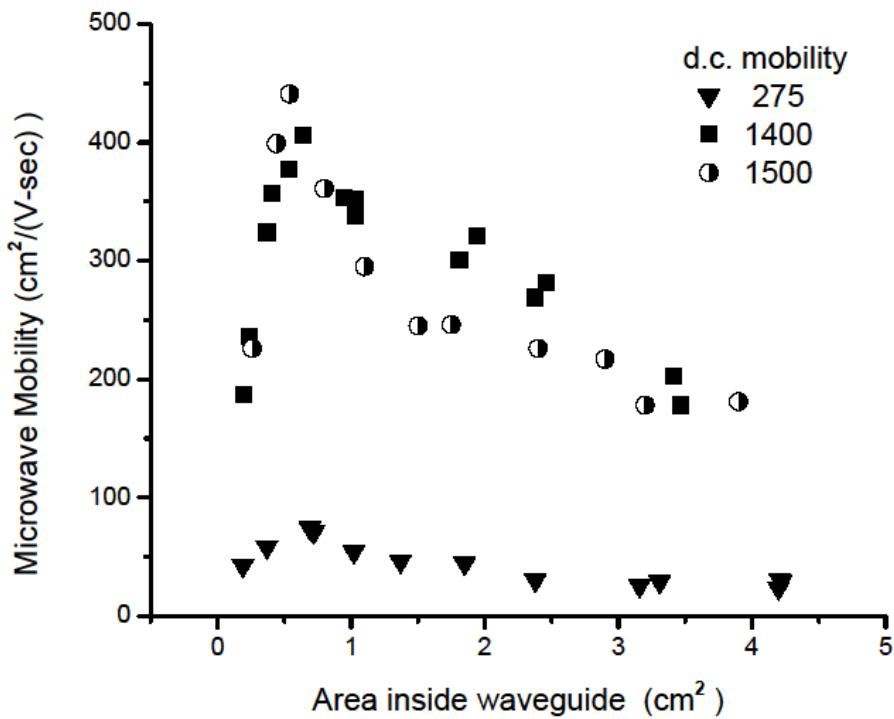

Figure 5. Measured microwave mobility for rectangular samples. The measured microwave mobility depends strongly on the area of the sample. The units of the mobility in the legend are $\frac{cm^2}{V-sec}$.

Thus, we can conclude that the measured microwave mobility and thus the calibration factor $\kappa$ for the Hall effect is highly geometry-dependent. We attempt to find this dependence by plotting the measured microwave mobility for only rectangular samples whose area inside the waveguide is greater than 0.52 cm². In fig. 6, we plot the microwave mobility as a function of width for samples of a given height. and in fig. 7, we plot the microwave mobility as a function of height for a given width. The height is measured perpendicular to the opening of the third arm of the series tee, and the width is measured parallel to the opening. In both figures, the measured mobility decreases as the samples increases in size. We interpret this to mean that as the edges of the sample become farther from the opening of the third arm of the tee, the measured signal decreases.

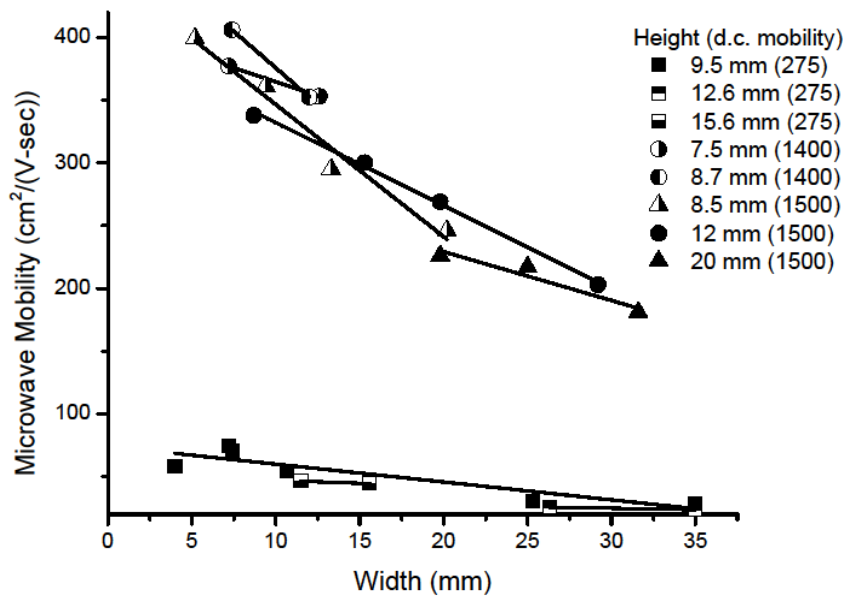

Figure 6. Measured microwave mobility as a function of width for samples of a given height. The height is measured perpendicular to the opening of the third arm of the series tee, and the width is measured parallel to the opening. The units of the mobility in the legend are $\frac{cm^2}{V-sec}$.

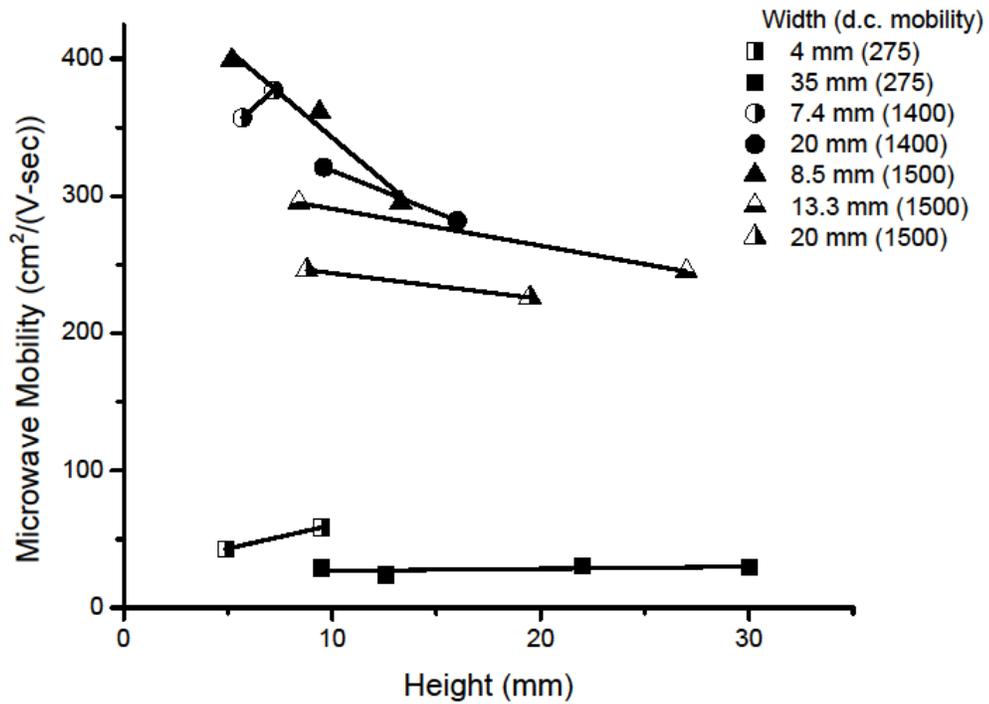

Figure 7. Measured microwave mobility as a function of height for samples of a given width. The height is measured perpendicular to the opening of the third arm of the series tee, and the width is measured parallel to the opening. The units of the mobility in the legend are $\frac{cm^2}{V-sec}$.

## 3. Microwave Resistivity

### A. Theory

The microwave Hall effect measures the mobility µ of a carrier. To determine the carrier concentration $n$ also requires measuring the conductivity, $\sigma = nq\mu$. The charge q has the magnitude of the electron's charge. To obtain the conductivity, we measure the sheet resistance of a wafer using a reflection method [13,14,15,16]. Our apparatus is shown in fig. 8.

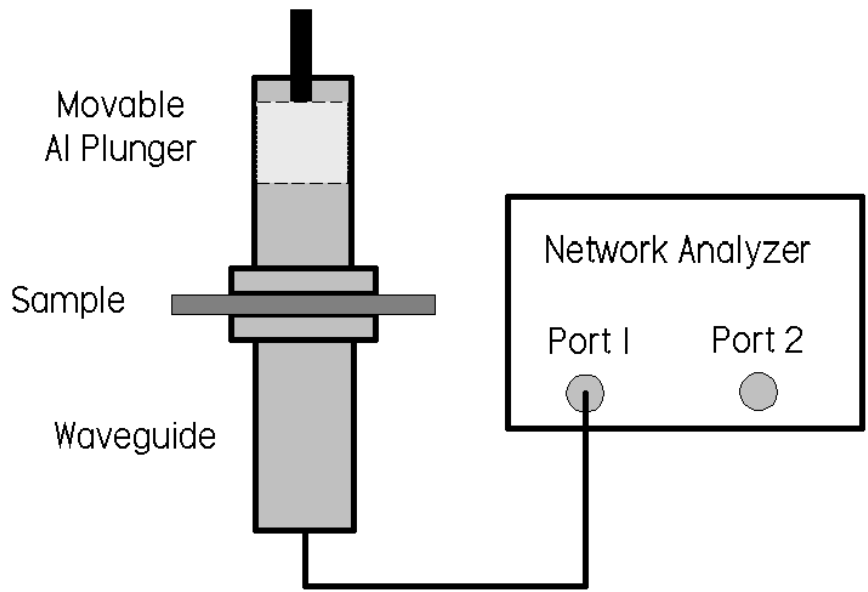

Figure 8. Schematic diagram of microwave resistivity apparatus.

We measure the microwaves reflected from a wafer at the end of an X band waveguide. The reflection coefficient, Γ, is related to the reactance $X$ and the resistance $R$ by

$$\Gamma = S_{11} = \frac{Z-Z_0}{Z+Z_0} = \frac{(R+iX)-Z_0}{(R+iX)+Z_0} \quad (3.1),$$

where the frequency dependent characteristic impedance of the WR90 waveguide is given by [20,21] where $f$ is the frequency in GHz:

$$Z_0 = \frac{377\ ohms}{\sqrt{1-(\frac{6.557\ GHz}{f})^2}} \quad (3.2).$$

We adjust the length of a shorted stub located behind the wafer to remove any reactive component in the impedance [22, 23]. We model the wafer as a resistance R (real component) in parallel with a reactance X (imaginary component). The length of the shorting stub is adjusted to add negative reactance in parallel with the wafer to cancel the wafer's reactance [22, 23]. Fig. 9 shows that the minimum value of the magnitude of $\Gamma$ corresponds to zero total reactance for the system.

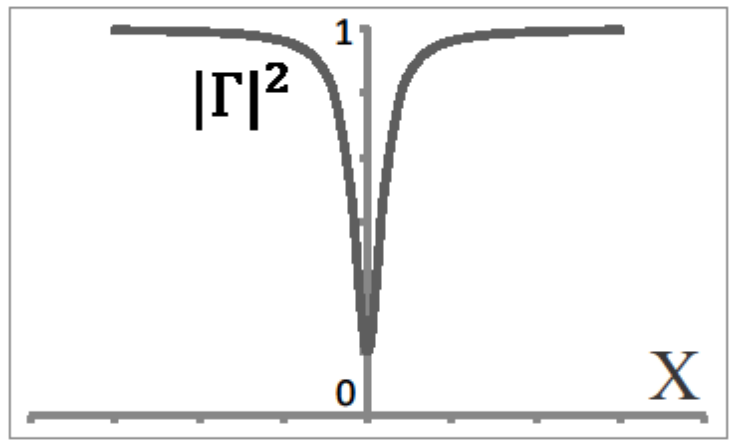

Figure 9. When the magnitude of the reflection coefficient $\Gamma = S_{11}$ is a minimum, the reactance $X$ of the sample plus waveguide system has been reduced to zero (see eq. 3.1).

We need to relate the equivalent resistance R of the wafer to its conductivity, since reflection method measures a resistance R. We can calculate the incident power $P_{inc}$:

$$P_{inc} = \frac{V^2}{Z_0} \quad (3.3)$$

where V is the voltage amplitude of the microwaves. The power absorbed $P_{abs}$ in the wafer:

$$P_{abs} = \frac{V^2}{R} \quad (3.4).$$

The formula for the reflection coefficient only involves the ratio

$$\frac{R}{Z_0} = \frac{P_{inc}}{P_{abs}} \quad (3.5)$$

and the voltage V drops out and thus does not need to be calculated.

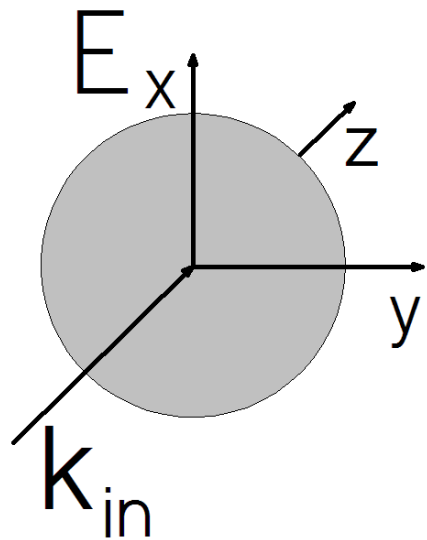

Figure 10. Geometry of the microwave measurement of the resistivity. An incident microwave field of amplitude $E_x$ and direction $k_{in}$ propagates along the z-axis into a semiconductor wafer in the x-y

plane. The incident microwave field induces currents in the wafer in the x direction that cause a reflection opposite to $\boldsymbol{k}_{in}$.

Assume that the microwaves are incident on a wafer of thickness $T$ perpendicular to the plane of the wafer (fig. 10). The electric field in eq. 3.6 is the electric field of the $TE_{10}$ mode [24] that is attenuated in the z direction (into the wafer):

$$E_x = E_{x0} e^{-z/\delta} \sin\left(\frac{\pi y}{L_y}\right) \quad (3.6),$$

and $E_x$ is parallel to the surface of the wafer. The waveguide is assumed to have its short axis along the x-axis. The current induced in the wafer is $\boldsymbol{J} = \sigma \boldsymbol{E}$. The absorbed power in the whole wafer is the integral over the volume of $\boldsymbol{J} \cdot \boldsymbol{E}$ in each elementary volume $d\tau$:

$$P_{abs} =$$
$$\int \sigma E^2 \, d\tau = \int \sigma E^2 \, dz \, dA = \int \sigma \left[ E_{x0} e^{-z/\delta} \sin\left(\frac{\pi y}{L_y}\right) \right]^2 dA \, dz \quad (3.7).$$

The incident power is given by the Poynting vector:

$$P_{inc} = \int \boldsymbol{E} \cdot \boldsymbol{H} \, dA = \int \frac{E^2}{Z_0} dA = \int \frac{\left[ E_{x0} \sin\left(\frac{\pi y}{L_y}\right) \right]^2}{Z_0} dA \quad (3.8).$$

We then substitute eq. 3.7 and eq. 3.8 into eq. 3.5 and divide to get:

$$R = \frac{1}{\sigma \int_0^T e^{-2z/\delta} dz} = \frac{1}{\sigma \delta \left( \frac{1 - e^{-2T/\delta}}{2} \right)} \quad (3.9).$$

If the sample is non-conductive, then the thickness is less than the skin depth, $T < \delta$, and the exponential becomes $e^{-2T/\delta} \approx 1 - 2T/\delta$, so that

$$R = \frac{1}{\sigma T} \quad (3.10).$$

If the sample is conductive, $T > \delta$, and the exponential drops out and

$$R = \frac{2}{\sigma \delta} \quad (3.11).$$

We calibrated the measurement of the reflection coefficient $S_{11}$ by using a highly conductive and highly polished silicon wafer. We found that a highly conductive ($\rho = 0.003 \; \Omega \cdot cm$) and highly polished silicon wafer reflected more microwave power than an ordinary unpolished copper block, despite the fact that the silicon wafer should have a sheet resistance about 45 times that of copper. This may be because the surface of the copper was rough and oxidized and this reduced the amount of microwave power that was reflected.

We measured $S_{11}$ of the sample relative to the $S_{11}$ of the silicon wafer, which was assumed to be unity. The frequency dependent characteristic impedance of the waveguide is given by eq. 3.2. We then calculated $\frac{Z}{Z_0}$ from $S_{11}$ and multiplied by $Z_0$ to find Z(=R).

B. Experimental Setup and Measurement Procedure

The apparatus is shown in Fig. 8. It consists of the sample under test sandwiched between two sections of waveguide, the end flanges of which are in contact with the wafer. A signal of the same frequency used for the microwave Hall measurement (about 9.5 GHz) is reflected off the sample and collected. We find that more accurate measurements are obtained if the frequency is swept through a small band (9.2-9.8 GHz) and the resulting $S_{11}$ is smoothed across the frequencies.

The calibration standard is placed in the apparatus and the magnitude and phase of the $S_{11}$ calibration are noted. With the sample under test in the apparatus, the shorting stub is adjusted until $S_{11}$ reaches a minimum in magnitude. The magnitude of $|S_{11}|$ and its phase $\theta$ relative to that of the calibration standard are noted, where the calibration standard is arbitrarily defined to have zero phase.

### C. Measurements

The reflection coefficient is calculated as: $\Gamma = \frac{|S_{11,sample}| \cdot \cos(\theta)}{|S_{11,calib}|}$. The resistance is calculated from: $R = Z_0 \cdot \frac{1+\Gamma}{1-\Gamma}$. Since the skin depth is greater than the wafer thickness for all samples tested, we may use the approximation $\rho = R \cdot T$.

The microwave resistivity method gives results within about 10% of the values obtained by the standard van der Pauw method, as shown in Table 1.

The results of the microwave resistivity measurement are with 4% of the resistivity measured using the van der Pauw method. This demonstrates that for the Si and Ge samples that we measured, the difference between the resistivity at d. c. and 10 GHz is small. We infer from this that the Hall mobility at 10 GHz is also close to the d. c. value.

| Material | Thickness [cm] | Skin Depth [cm] | $\rho_{DC}$ [$\Omega \cdot cm$] | $\rho_{MW}$ [$\Omega \cdot cm$] |
|---|---|---|---|---|
| Si:P (2-in) | 0.029 | 0.11 | $4.67 \pm 0.06$ | $4.5 \pm 0.2$ |
| Si:P (3-in) | 0.052 | 0.09 | $3.30 \pm 0.03$ | $3.8 \pm 0.2$ |
| Si:P (3-in) | 0.052 | 0.10 | $3.53 \pm 0.03$ | $3.6 \pm 0.1$ |
|  |  |  |  |  |
| Si:B (2-in) | 0.028 | 0.06 | $1.45 \pm 0.04$ | $1.6 \pm 0.1$ |
| Si:B (3-in) | 0.038 | 0.15 | $8.09 \pm 0.06$ | $8.1 \pm 0.3$ |
|  |  |  |  |  |
| Ge (2-in) (#834B) | 0.041 | 0.3 | $34.5 \pm 0.3$ | $40 \pm 2$ |
| Ge (2-in) (#1053A) | 0.047 | 0.08 | $2.36 \pm 0.02$ | $2.5 \pm 0.2$ |
| Ge (2-in) (#1310A) | 0.042 | 0.19 | $14.3 \pm 0.1$ | $16 \pm 2$ |

Table 1. Results for microwave resistivity measurements in which the $S_{11}$ signal is smoothed over the same frequency range from 9.2-9.8 GHz that was used for the Hall measurement and the shorting stub is adjusted to minimize the smoothed signal. $\rho_{MW}$ is the microwave resistivity and $\rho_{DC}$ is the van der Pauw resistivity.

### D. Extensions of the method

This method reliably determines the resistance of a sample, but a determination of the resistivity depends on knowledge of the skin depth, which itself depends on the resistivity. Thus, the resistivity can be reliably determined only for samples whose skin depth is comparable to or greater than their thickness, so that the skin depth is practically the entire wafer.  Once a resistivity is measured, self-consistency is checked by calculating the skin depth for that resistivity and comparing it to the sample thickness.

The apparatus has thus far been tested on two and three inch wafers, which are large enough to fully cover the waveguide aperture. Smaller samples can be mounted on a very thin glass slide to minimize power losses through the gap between waveguide flanges.

We believe that the method will work on thin film samples, and linear Hall data has been obtained in preliminary tests on a GaInAs film. It has the advantage of not being affected by holes in a film which would cause errors in van der Pauw measurements due to violation of the requirement that the sample be simply connected.

## 4. Conclusion

We measured the Hall effect in silicon and germanium circular wafers in a microwave series tee that is part of a microwave interferometer.  We used an X band network analyzer to measure the Hall signal after we null out a large offset signal at zero magnetic field. The application of a magnetic field of up to 1.5 tesla caused the Hall effect signal to increase linearly with the field.  The phase of the Hall signal changed sign when the magnetic field was reversed.  The microwave interferometer is needed to increase the sensitivity of the method by removing the large offset signal which would otherwise completely mask the Hall effect.

We measured the resistivity of the same wafers using a microwave reflection method.

The microwave Hall signal depended on the size of the sample and the position of the sample in the microwave tee.  We calibrated out the geometry effects by comparing the microwave measurements with the d. c. mobility and resistivity found from the van der Pauw method.

We also studied the dependence of the microwave Hall signal on the shape of the sample for rectangular samples.